\documentclass[aps,prc,twocolumn,superscriptaddress,showpacs]{revtex4-1}

\usepackage{graphicx,amsmath,amssymb,bm}
\usepackage{rotating}

\begin{document}

\title{Beyond the Neutron Drip-Line:
The Unbound Oxygen Isotopes $^{25}$O and $^{26}$O}

\author{C.~Caesar}
\affiliation{Institut f\"ur Kernphysik, Technische Universit\"at Darmstadt, 64289 Darmstadt, Germany}
\affiliation{GSI Helmholtzzentrum f\"ur Schwerionenforschung, D-64291 Darmstadt, Germany}
\author{J.~Simonis}
\affiliation{Institut f\"ur Kernphysik, Technische Universit\"at Darmstadt, 64289 Darmstadt, Germany} 
\affiliation{ExtreMe Matter Institute EMMI, GSI Helmholtzzentrum f\"ur Schwerionenforschung GmbH, 64291 Darmstadt, Germany}
\author{T.~Adachi}
\affiliation{KVI, University of Groningen, Zernikelaan 25, NL-9747 AA Groningen, The Netherlands}
\author{Y.~Aksyutina}
\affiliation{GSI Helmholtzzentrum f\"ur Schwerionenforschung, D-64291 Darmstadt, Germany}
\affiliation{ExtreMe Matter Institute EMMI, GSI Helmholtzzentrum f\"ur Schwerionenforschung GmbH, 64291 Darmstadt, Germany}
\author{J.~Alcantara}
\affiliation{Departamento de F\'{i}sica de Part\'{i}culas, Universidade de Santiago de Compostela, 15706 Santiago de Compostela, Spain}
\author{S.~Altstadt}
\affiliation{Goethe-Universit\"at Frankfurt am Main, 60438 Frankfurt am Main, Germany}
\author{H.~Alvarez-Pol}
\affiliation{Departamento de F\'{i}sica de Part\'{i}culas, Universidade de Santiago de Compostela, 15706 Santiago de Compostela, Spain}
\author{N.~Ashwood}
\affiliation{School of Physics and Astronomy, University of Birmingham, Birmingham B15 2TT, United Kingdom}
\author{T.~Aumann}
\email[E-mail:~]{t.aumann@gsi.de}
\affiliation{Institut f\"ur Kernphysik, Technische Universit\"at Darmstadt, 64289 Darmstadt, Germany}
\affiliation{GSI Helmholtzzentrum f\"ur Schwerionenforschung, D-64291 Darmstadt, Germany}
\author{V.~Avdeichikov}
\affiliation{Department of Physics, Lund University, S-22100 Lund, Sweden}
% ...

\collaboration{R3B collaboration}
\noaffiliation

\date{\today}

\begin{abstract}
The very neutron-rich oxygen isotopes $^{25}$O and $^{26}$O are
investigated experimentally and theoretically.  The unbound states 
are populated in an experiment performed at the R3B-LAND setup at GSI via
proton-knockout reactions from $^{26}$F and $^{27}$F at relativistic
energies around 442~MeV/nucleon and 414~MeV/nucleon, respectively. 
From the kinematically complete
measurement of the decay into $^{24}$O plus one or two neutrons, the
$^{25}$O ground-state energy and width are determined,
and upper limits for the $^{26}$O ground-state energy and lifetime
are extracted. In addition, the results provide indications for
an excited state in $^{26}$O at around 4~MeV. The
experimental findings are compared to theoretical shell-model
calculations based on chiral two and three-nucleon (3N) forces,
including for the first time residual 3N forces, which are shown to
be amplified as valence neutrons are added.
\end{abstract}


\maketitle

\section{INTRODUCTION}

Understanding the properties of nuclei with extreme neutron-to-proton
ratios presents a major challenge for rare-isotope beam experiments
and nuclear theory. Nuclei located at and beyond the neutron dripline
play a crucial role in this endeavor. Experimentally, the neutron
dripline has been established up to oxygen~\cite{lang85,guil90,tara97} with $^{24}$O
being the last bound isotope, while it extends considerably further in
fluorine~\cite{saku99}.
Recently, it has been shown that the anomalous behavior in
the oxygen isotopes is due to the impact of three-nucleon (3N) forces, which provide
repulsive contributions to the interactions of valence
neutrons~\cite{otsu10}, connecting the frontier of neutron-rich
nuclei to the theoretical developments of nuclear forces.

Another striking feature in the oxygen isotopic chain is the doubly-magic nature of
$^{22}$O and $^{24}$O~\cite{hoff08,kanu09,tsho12,thir00,brow05,bech06} 
in strong contrast to the
lighter elements, where the dripline is marked by nuclei exhibiting a
loosely-bound halo structure.
The neutron-rich oxygen isotopes also provide interesting insights, when
viewed coming from their stable isotones. 
As protons are removed, 
the attractive contribution from the proton-neutron tensor
force decreases, thus, opening up the $N=16$ neutron shell gap for
oxygen~\cite{otsu05}, while reducing the gap at $N=20$ 
which is very prominent in stable nuclei.

How the structure evolves beyond $^{24}$O towards $N=20$ is thus
of central interest. 
Currently, $^{25}$O and $^{26}$O are at the limit of experimental availability. For the 
former isotope, the ground-state resonance energy and width have been reported~\cite{hoff08}.
For the latter, its position has been measured previously~\cite{lund12}.
Taking advantage of the large angular acceptance for neutrons in the 
R3B-LAND experiment~\cite{auma07,R3BTDR}, we investigate the unbound
isotopes $^{25}$O and $^{26}$O in an extended energy range with an essentially 
constant efficiency up to a decay energy of 4~MeV and 8~MeV, respectively.

It has been speculated that the unbound isotopes $^{26}$O and $^{28}$O might have
a rather long lifetime, which would constitute the first example of
neutron radioactivity~\cite{grig11}. Our present result establishes an upper
limit for the lifetime of the $^{26}$O ground state. We then combine the
experimental investigation with theoretical calculations based on
chiral two-nucleon (NN) and 3N forces, where we focus on the
increasing contribution from residual three-neutron forces as
neutrons are added.

\section{EXPERIMENT}

The experiment was carried out at the GSI Helmholtzzentrum in
Darmstadt using the R3B-LAND reaction setup. Beams of light
neutron-rich nuclei were produced by fragmentation reactions of a
490~MeV/nucleon $^{40}$Ar primary beam in a 4~g/cm$^2$ Be target. Ions
with a magnetic rigidity of 9.88($\pm1$\%)~Tm corresponding to an 
$A/Z$ ratio of about 3 were selected by the fragment separator 
FRS~\cite{geis92} 
and transported to the experimental area. Energy-loss and time-of-flight
measurements allowed for the identification of the incoming ions on an
event-by-event basis. The beam cocktail contained $^{26,27}$F ions ($\sim$1\%),
which were selected to populate the unbound
states in $^{25,26}$O  \textit{via} one-proton knockout reactions. 
The energies (intensities) of the $^{26}$F and $^{27}$F beams 
were 442~MeV/nucleon and 414~MeV/nucleon (1~Hz and 0.3~Hz), respectively.
Different secondary targets 
(922~mg/cm$^{2}$ CH$_{2}$, 935~mg/cm$^{2}$ C, and  2145 ~mg/cm$^{2}$ Pb) 
were used, and all shown spectra display
the contributions from all targets. 
The target was surrounded by the 4$\pi$ Crystal Ball detector~\cite{meta83} 
consisting of 160 NaI crystals for detecting photons and
light particles emitted at laboratory angles larger than $\pm7^{\circ}$
relative to the beam axis. Position and energy loss of
the beam and fragments behind the target were measured by two silicon-strip 
detectors before deflection in the large-gap dipole magnet
ALADIN. Two further position measurements behind the magnet using
scintillating fiber detectors  \cite{cub98,maha09} allowed for tracking of the ions through
the dipole field. Together with time-of-flight and energy-loss
measurements, this provides the magnetic rigidity and atomic number, and thus the mass of
the fragments. 

Neutrons from the decay of unbound states were detected
at a distance of around 12~m downstream of the target by the LAND neutron
detector \cite{blai92} with an efficiency of 92\% for single neutrons and
with an angular acceptance of $\pm$79~mrad around the beam axis. A
similar experimental setup and analysis scheme is described in
Ref.~\cite{pali03} in more detail.

\section{ANALYSIS AND RESULTS}
\subsection{$^{25}$O ground-state resonance}

\begin{figure}
\includegraphics[clip,keepaspectratio,width=1.0\columnwidth]{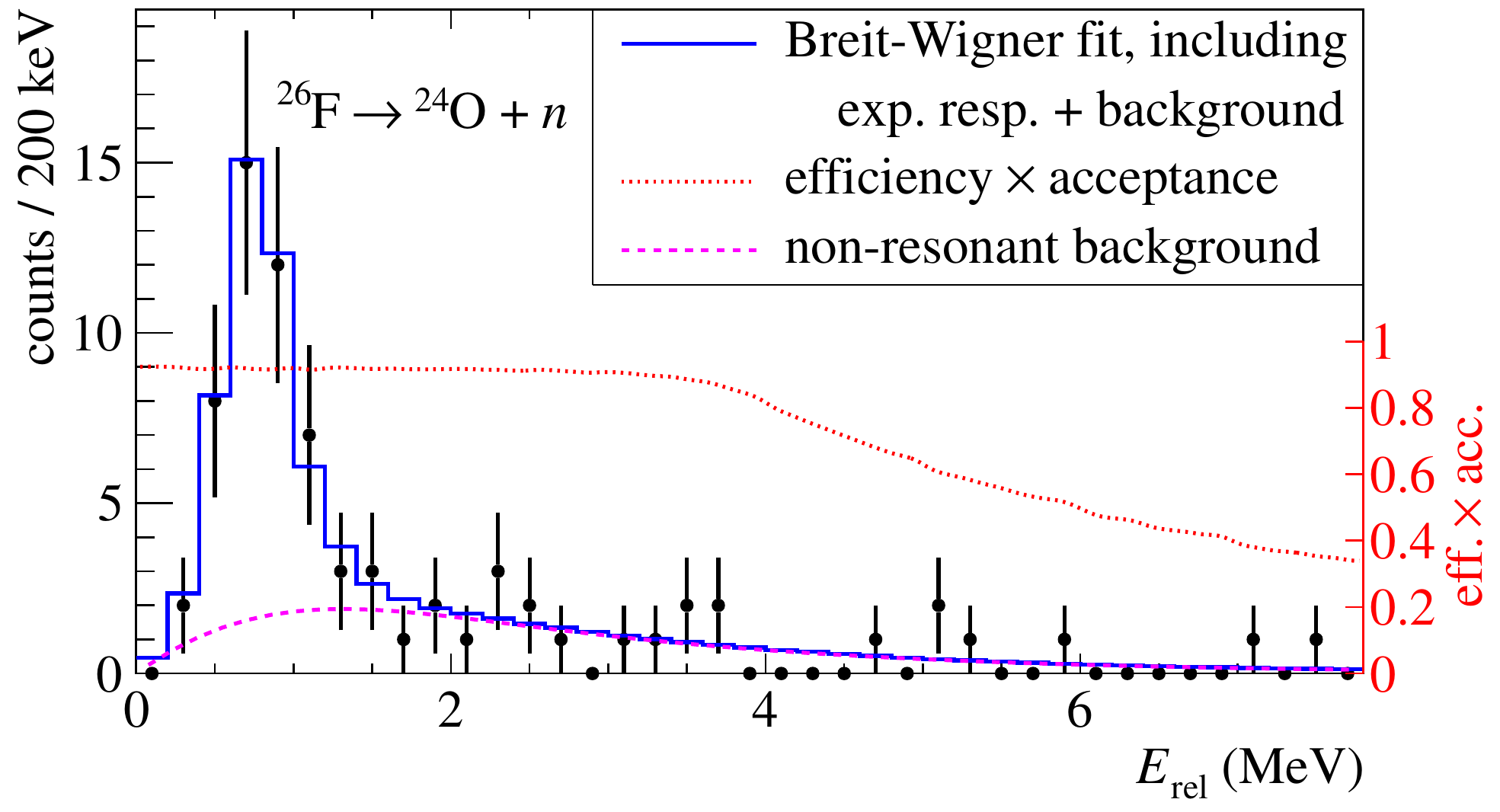}
\caption{\label{fig:25O} (color online) Relative-energy spectrum 
of the $^{24}$O$+n$ system measured in a proton-knockout reaction
from $^{26}$F. The blue solid line shows a Breit-Wigner fit to the 
data, which includes the experimental response and a non-resonant 
background (purple dashed curve). The red dotted curve 
indicates the experimental efficiency including the acceptance.
}
\end{figure}

From the measurements of the momenta of outgoing fragments and
neutrons, the two and three-body relative-energy ($E_{\mathrm{rel}}$) spectra are
reconstructed for one and two-neutron events. Fig.~\ref{fig:25O}
shows the $^{24}$O+$n$ $E_{\mathrm{rel}}$ spectrum after proton removal
from $^{26}$F. A prominent peak structure is visible at about 700~keV
corresponding to the ground-state resonance of $^{25}$O. The position $E_{r}$
and width $\Gamma$ of the resonance have been extracted by fitting a
Breit-Wigner distribution with an energy-dependent width using the 
following function~\cite{lane58}
\begin{equation}
 \label{eq_BreitWigner}
f(E;E_{r},\Gamma) = \dfrac{\Gamma}{(E_{r}+\Delta-E)^2 + \Gamma^2  / 4 } \; ,
\end{equation}
with the resonance shift $\Delta$ set to zero and the width given by
$\Gamma = 2P_{l}(E;R) \times \gamma^2$ with the reduced width amplitude $\gamma$
and the penetration factor $P_{l}$. The penetration 
factor (taken from Ref.~\cite{bohr69}) depends on the channel radius $R$,
the energy $E$ and angular momentum $l$. As the distribution was found to be 
insensitive to changes in $R$ between 3.5 and 6~fm, a channel radius of 
4~fm was used. An angular momentum of $l=2$ is used, since the additional 
neutron of $^{25}$O compared to $^{24}$O is most likely in the $0d_{3/2}$ orbital.

This distribution has been convoluted with the experimental response. A non-resonant 
background (BG) has been modeled as the product of an error and of an exponential 
function:
\begin{equation}
	\label{eq_background}
	f(E) =  a \times \mathrm{erf}(b E) \times e^{- c E} \; ,
\end{equation}
with free paramters $a$, $b$ and $c$. The sum of the 
convoluted Breit-Wigner and background functions was used to 
fit the experimental data.

\begin{figure}
	\includegraphics[clip=,keepaspectratio,width=1.0\columnwidth]{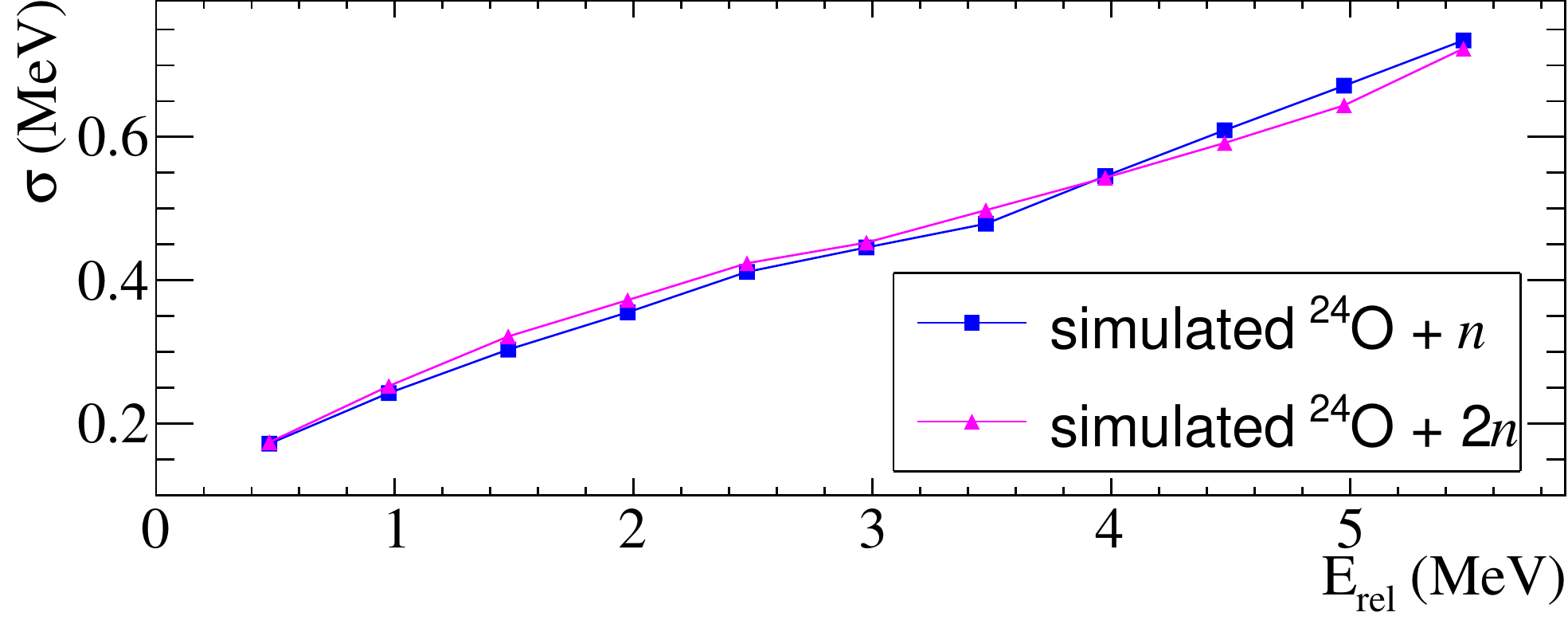}
	\caption{\label{fig:resolution}(color online) Resolution ($\sigma$) as a function of the relative energy ($E_{\mathrm{rel}}$). 
		The simulated response matrices, as shown in Fig.~\ref{fig:LEG} for the 2$n$ case, have been used to determine the 
		resolution for the individual $E_{rel}$ values.}
\end{figure}

\begin{figure}
\includegraphics[clip=,keepaspectratio,width=1.0\columnwidth]{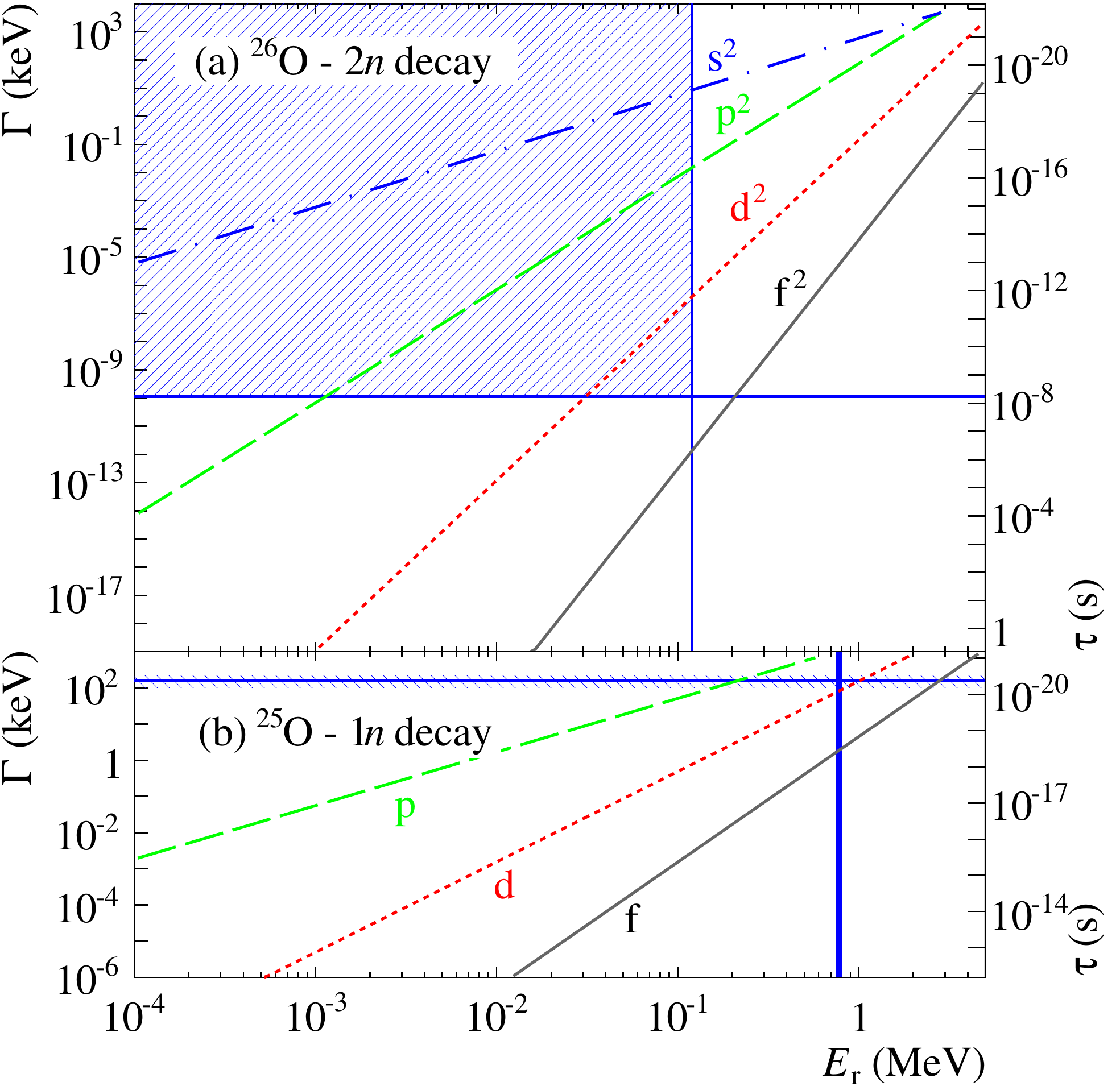}
\caption{\label{fig:grig} (color online) Width and lifetime as a 
function of resonance energy for $^{26}$O (upper panel) and $^{25}$O 
(lower panel). The curves show theoretical expectations for different
$l$-values of the neutron(s) from Ref.~\cite{grig11}. For $^{26}$O,
the (2$\sigma$) upper limits for the resonance energy and lifetime are given
by horizontal and vertical blue lines. The allowed region defined
by these limits is represented by the hatched area. For $^{25}$O,
the average of the present experiment and of the results from Hoffman \textit{et al.}~\cite{hoff08}
is given by the horizontal and vertical blue lines, with the line-thickness (hatched zone) corresponding to 2$\sigma$ errors.}
\end{figure}

For the $\chi^2$ minimization procedure, Pearson's
chi-square method~\cite{bake84}, using errors of the parent
distribution according to a Poisson probability distribution, has been
used, as the usual method with errors estimated from the number of
counts gives inaccurate results in case of low statistics. The extracted 
position (width) of $E_r = 725 ^{+54}_{-29}$~keV ($\Gamma = 20 ^{+60}_{-20}$~keV) 
is in  agreement with the previously reported value~\cite{hoff08}
within 1$\sigma$ (2$\sigma$), see Table~\ref{tab}. 
Our result is in agreement with a single-particle width $\Gamma_{sp} = 65$~keV
calculated 
for a pure $d$-state.
The relatively large experimental
error on the width is due to the instrumental energy resolution which
dominates the apparent width, see Fig.~\ref{fig:resolution}.
For further discussion, results from literature and the present result were averaged 
according to Ref.~\cite{barl04}  resulting in
$E_r =768 ^{+19}_{-9}$~keV and $\Gamma = 160 ^{+30}_{-30}$. 
These values are compared in the lower panel of Fig.~\ref{fig:grig} 
to the expected widths and
lifetimes as a function of resonance energy for different neutron
angular momenta $l$ (adopted from Fig. 2 (b) of Ref.~\cite{grig11}). 
We note that the averaged width is close to the estimated value
for a $d$-state as given in Ref.~\cite{grig11}.

\subsection{$^{26}$O ground-state resonance}

\begin{figure}
\includegraphics[clip=,keepaspectratio,width=1.0\columnwidth]{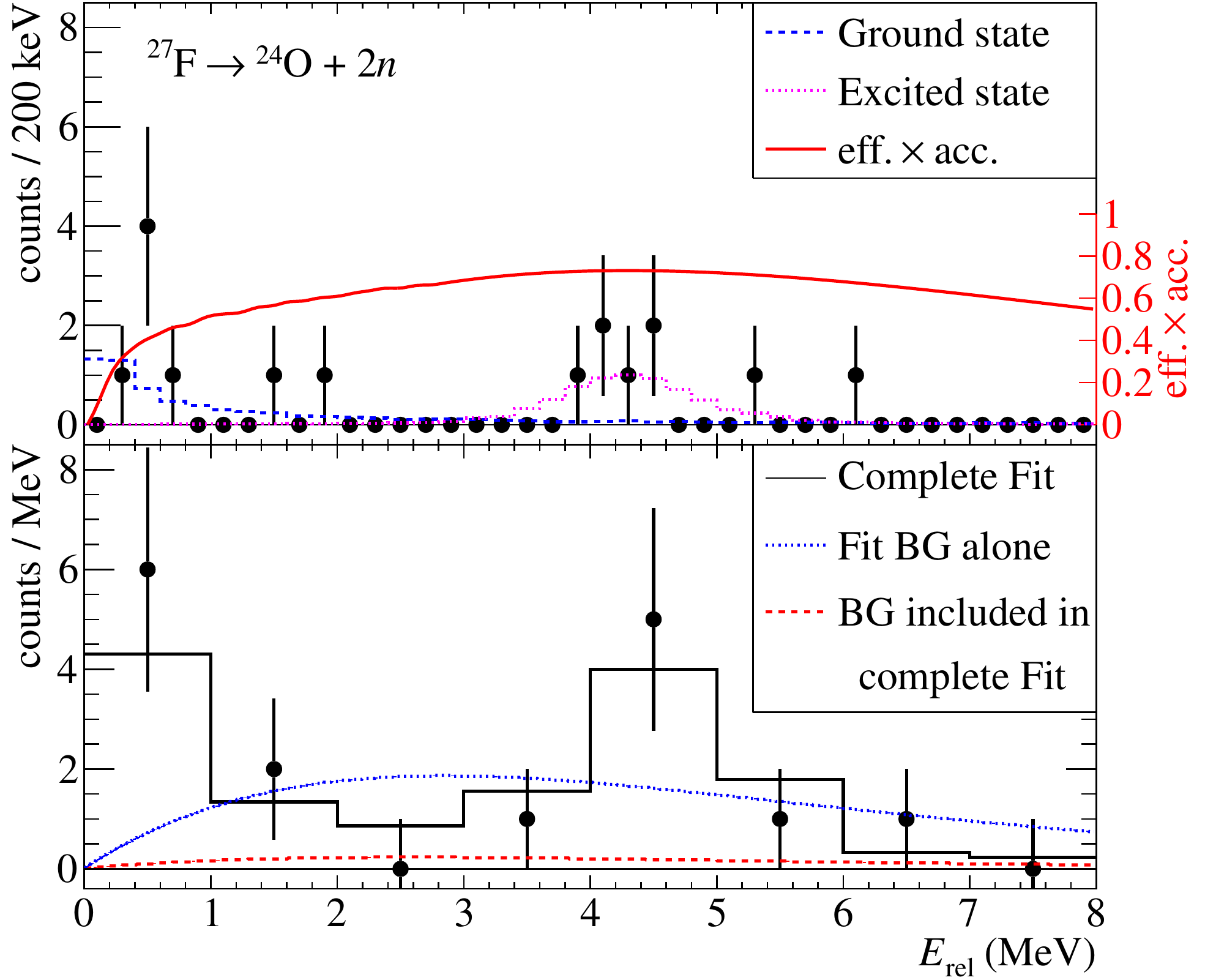}
\caption{\label{fig:26O} (color online) Two-neutron channel
measured in coincidence with $^{24}$O after one-proton removal from 
$^{27}$F. The symbols with error bars represent the measured 
$^{24}$O+$2n$ three-body relative-energy spectrum. Top panel:
the red curve displays the experimental detection efficiency including acceptance
for the $2n$ channel; the dotted and dashed histogram shows the most
probable fit to the data including two states in $^{26}$O
at $25\pm25$~keV and around 4~MeV. Bottom panel:
the symbols show the same data with a 1-MeV bin size;
the black solid line displays a fit to the data including two resonances and
a background (red dashed curve); the blue dotted curve shows a fit using 
only a non-resonant background (see text for details).}
\end{figure}

\begin{figure}
\includegraphics[width=1.0\columnwidth]{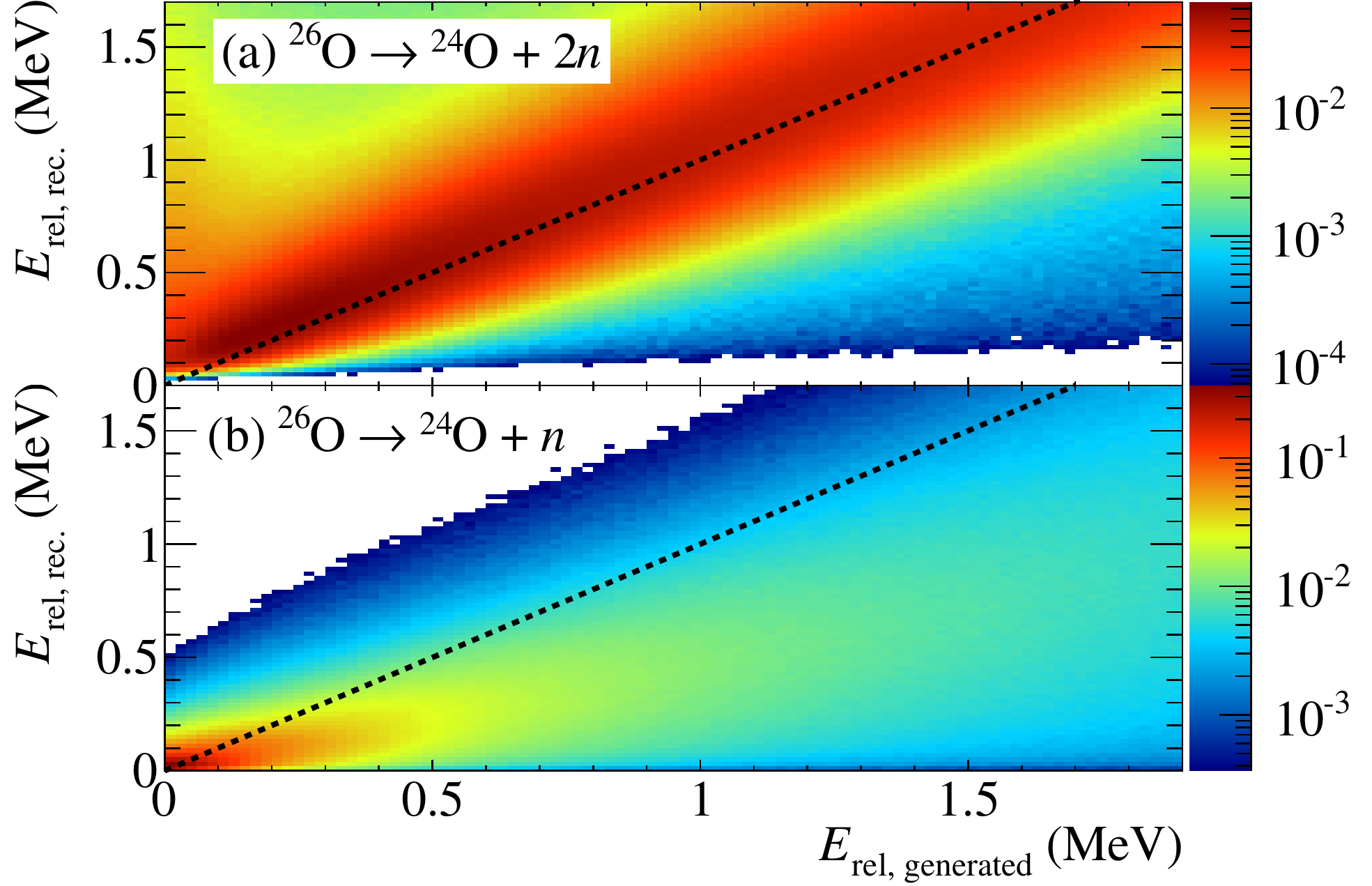}
\caption{\label{fig:LEG} (color online) Simulated instrumental response for
detection of a $2n$ decay of $^{26}$O with relative energy $E_{\mathrm{rel}}$.
The upper panel shows the reconstructed energy $E_{\mathrm{rel, rec.}}$ for two detected neutrons
in coincidence with $^{24}$O, while the lower part shows the
$^{24}$O+$n$ relative-energy spectrum for events, in which 
only one neutron has been detected due to the limited efficiency or acceptance of the 
detector or due to the multi-neutron recognition efficiency.
$E_{\mathrm{rel, generated}}$ refers to the initial energy  
used as input to simulate the decay and $E_{\mathrm{rel, rec.}}$ 
is the reconstructed energy after simulation and analysis.}
\end{figure}

The experimental $E_{\mathrm{rel}}$ spectrum for $^{26}$O is shown in
Fig.~\ref{fig:26O}, where $^{24}$O
has been detected in coincidence with two neutrons. Two groups of
events are observed: below 1~MeV and around 4~MeV. The 
experimental response, indicated by the red curve, is rather constant
over the displayed energy region, but exhibits a steep fall-off for
energies below 500~keV. For such small relative energies, the neutrons
are not well separated in space and time when interacting in the detector and
can thus hardly be distinguished from $1n$ events. The effect can be
seen quantitatively in the two-dimensional response matrices shown in
Fig.~\ref{fig:LEG}. The energy reconstructed from the simulation
is plotted versus the generated one in the upper panel, showing 
a band along the diagonal with a width reflecting the instrumental 
resolution, which is shown in Fig.~\ref{fig:resolution}.  
For low  generated relative energies ($\lesssim$100~keV), 
the events spread to a higher reconstructed energy and are, in addition, 
reconstructed as 
$1n$ events with a large probability. 
This can be seen in the lower panel of Fig.~\ref{fig:LEG}, 
which shows the reconstructed $^{24}$O+$n$ relative energy
spectrum for the events falsely identified as $1n$ events (either due to the effect 
discussed above at low relative energies, or due to limited coverage of the detector
for high relative energy).
The simulation is based on measured real
$1n$ events from deuteron breakup reactions. The $2n$ events are
generated by overlaying the shower patterns of secondary particles from these measured $1n$
events according to the simulated positions on the neutron detector 
and are then analyzed in the same way as the experimental
data \cite{bore03}. The generated response matrices thus do not rely on a simulation
of the reactions in the neutron detection process.

\begin{figure}
\includegraphics[width=\columnwidth]{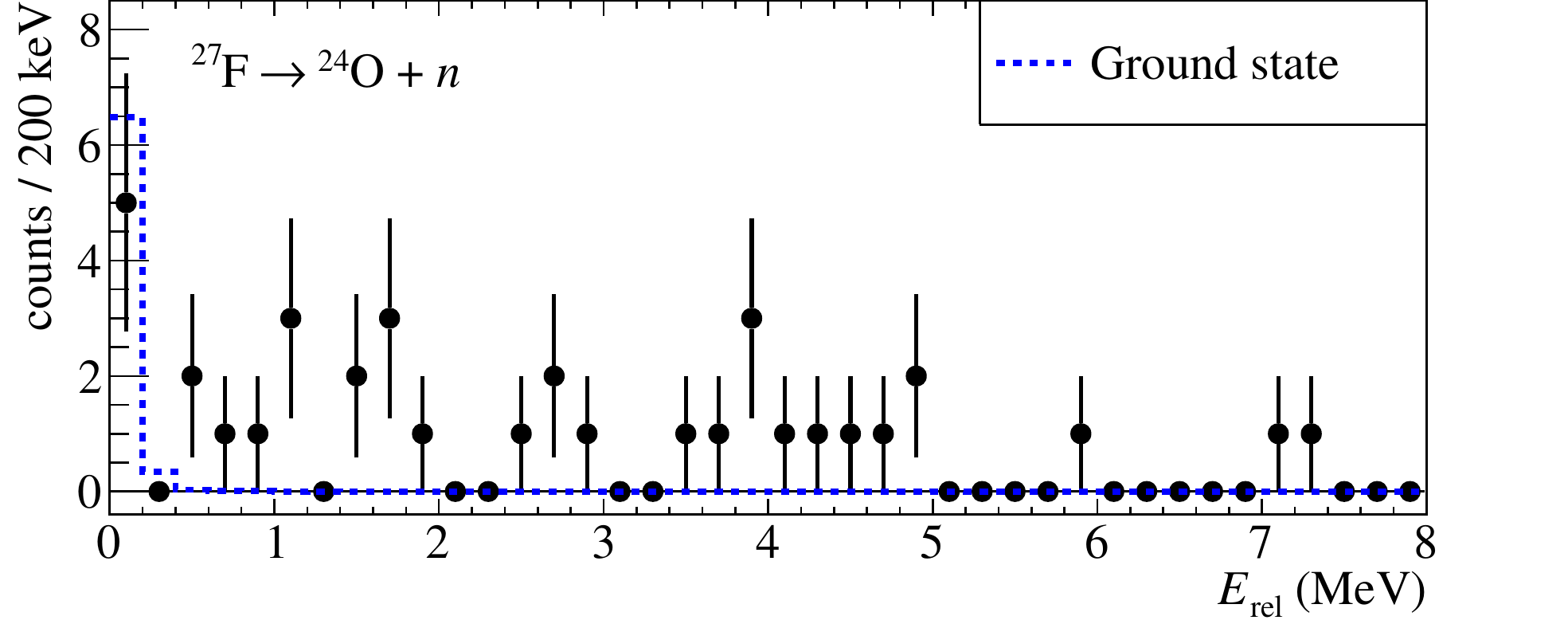}
\caption{\label{fig:26O_1n} (color online) One-neutron channel
measured in coincidence with $^{24}$O after one-proton removal from 
$^{27}$F. The data points represent the measured 
$^{24}$O+$1n$ two-body relative-energy spectrum.
The blue dashed curve displays the 1n contribution of the 
most probable fit to the $1n$ and $2n$ data for the $^{26}$O ground-state
at $25\pm25$~keV, to which the 5 counts in the first bin at 100 keV are 
attributed (see text).}
\end{figure}

The effect discussed above is clearly visible in the $^{26}$O data of
Fig.~\ref{fig:26O} and Fig.~\ref{fig:26O_1n}.
 In this $1n$ channel 
(Fig.~\ref{fig:26O_1n}), the
accumulation of events at very low energy in the first bin is
compelling. This feature is not present in the $1n$ events measured in the
proton knockout from $^{26}$F, as seen in Fig.~\ref{fig:25O}.
The events in the first bin of the $^{24}$O+$n$ spectrum
are the characteristic fingerprint of a very low-lying state in
$^{26}$O. The events with energies above 0.2~MeV
can be attributed to several processes, such as a possible
direct two-nucleon knockout reaction.
In particular, the events between 0.2~MeV
and 2~MeV could result from $pn$ knockout from $^{27}$F populating 
the $^{25}$O ground-state resonance, shown in Fig.~\ref{fig:25O}.
At higher energies, $2n$-decay events can contribute due to the limited detector 
acceptance. According to the simulation, three counts are expected in the 
$1n$ spectrum between 0 and 5~MeV stemming from the resonance-like
structure in the $2n$ channel at 4~MeV. In addition, knockout of more
deeply bound protons is expected, yielding higher excitation energies in 
$^{26}$O, which will appear as a broad background in the $1n$ spectrum.

We have performed a simultaneous statistical analysis of $1n$ and $2n$
coincidences, starting with the hypothesis of a low-lying state in
$^{26}$O. The dotted histograms in Fig.~\ref{fig:26O} (upper panel) and 
Fig.~\ref{fig:26O_1n} display the most probable result
yielding a position for the $^{26}$O ground state of $25\pm25$~keV.
Again, the $\chi^2$ minimization using Poisson distributed
errors of the response function has been used. The response functions
have been used to approximate the line shape.

Another method~\cite{bevi69}, which is independent of the
binning of the experimental data, is to use the
measured relative energy in each event in conjunction with the transposed response
matrix to calculate the probability distribution of the true energy
(Bayesian approach with uniform prior). The resulting Bayesian
interval runs from 0 to 40/120~keV at a 68\%/95\% confidence level (c.l.).
Again, both the $1n$ and $2n$ data are considered simultaneously.

We cannot exclude a very small value
close to zero for position and width from the energy measurement alone, potentially leading to a rather
long-lived $^{26}$O ground state, which would constitute the first
case of neutron radioactivity as speculated in Ref.~\cite{grig11}: As
can be seen from Fig.~\ref{fig:grig}, the lifetime for a $d^2$-state
could reach seconds for a resonance position well below 1~keV. Such a long
lifetime, however, can be excluded from the fragment measurement. The
distance of the target to the middle of the dipole magnet measures
256~cm, corresponding to a flight time of 11.8~ns. If $^{26}$O
would decay after that time, a fragment mass greater than 24 would be
reconstructed by the tracking procedure. Also, no neutron coincidences 
should be observed in that case, since the fragment is bent by 7 degree 
after passing half of the dipole field. 

\begin{figure}
\includegraphics[clip=,keepaspectratio,width=1.0\columnwidth]{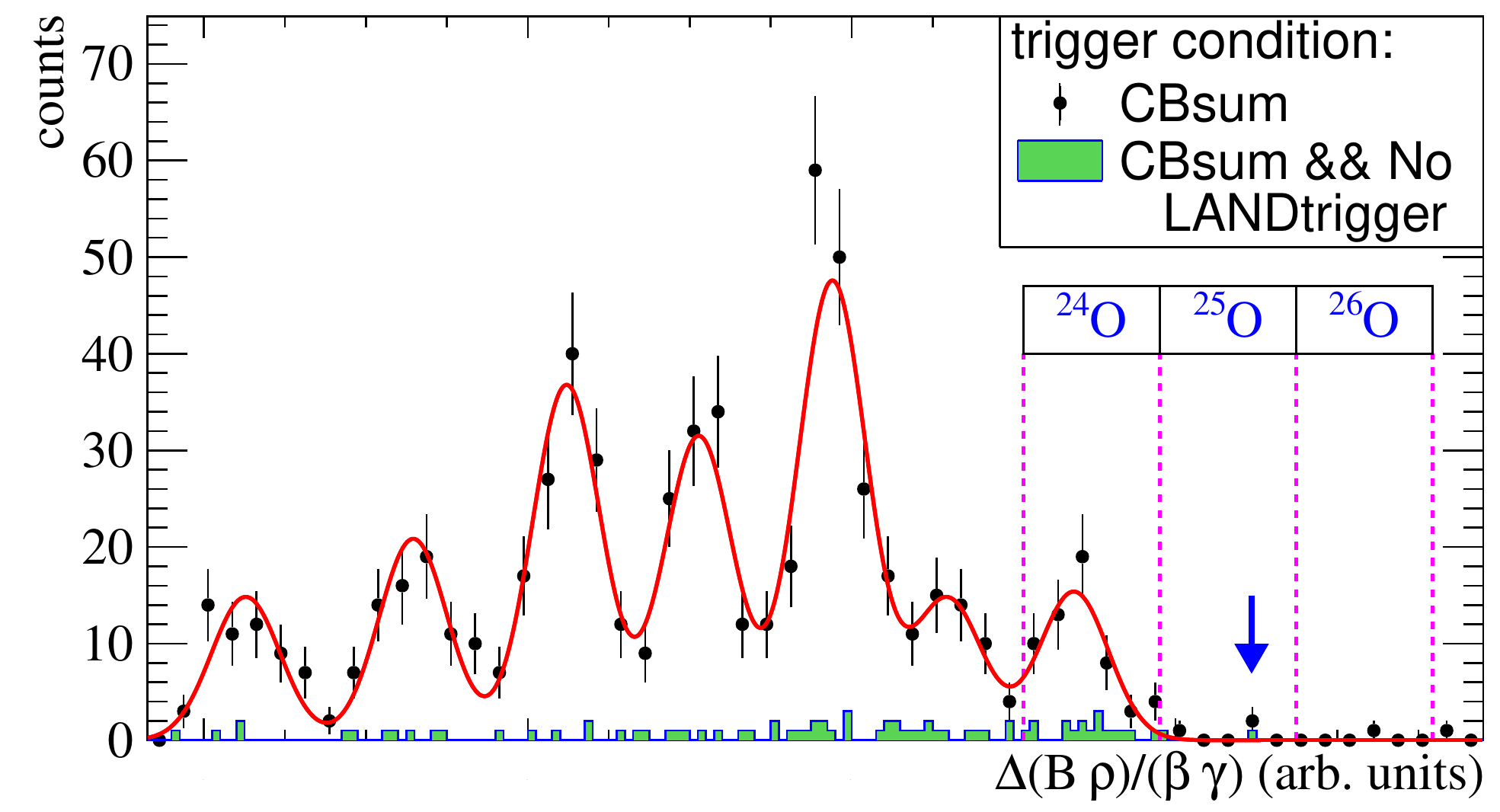}
\caption{\label{fig:Mass_26O}(color online) Fragment mass distribution for 
incoming $^{27}$F and outgoing oxygen isotopes ($Z=8$), 
obtained by tracking the fragment through the magnetic field. 
The gates for $^{24}$O, $^{25}$O and $^{26}$O 
are indicated by the vertical dashed lines. The spectrum 
generated by requiring the CB-sum trigger is shown
as the black data points. Applying the condition that 
no neutron is detected in coincidence, the green histogram remains,
In which one event is visible in the $^{25}$O oxygen 
gate (indicated by the blue arrow).
}
\end{figure}

In order to obtain an upper limit on the number of events belonging to the previously described
event class ($A > 24$ and no neutron in coincidence), the fragment mass spectrum has been 
inspected under the condition that no neutron is detected in coincidence, \textit{i.e.}, the neutron detector
was used as a veto (the efficiency to detect a neutron at forward direction is 92\%).
The reaction trigger was provided by the Crystal Ball (CB), since for the case of proton knockout reactions,
the knocked-out proton is detected at large angles with high probability in the CB. This is not only the case
for  $^{27}$F$(p,2p)^A$O quasi-free reactions on the hydrogen in the CH$_2$ target, but also for knockout 
reactions induced by composite targets.
The resulting fragment-mass distribution for 
incoming $^{27}$F and outgoing oxygen isotopes ($Z=8$) is compared in Fig.~\ref{fig:Mass_26O}
for the described trigger condition with the distribution obtained with neutron coincidences. 
As can be seen by comparing the two distributions in Fig.~\ref{fig:Mass_26O},
there are only very few events observed without coincident neutrons, which can be explained
by the neutron detection efficiency of 92\% and by a small amount of background events. 
Only one event appears above the $^{24}$O mass, which could be attributed to
background (since there are also few events in that mass range with neutron coincidences).
This one count, however, provides an upper limit on the survival probability of $^{26}$O. 
Assuming a Poisson distribution, an expectation value of 4.9 would yield a probability of 
$>95.5$\% (2$\sigma$)  to detect more than one event. 
From the analysis of the energy spectra discussed above (see Fig.~\ref{fig:26O}), the
number  $N(t=0)=20.5$ of produced $^{26}$O in the ground state can be estimated. 
From the initial number of
$^{26}$O, the upper limit $N(t=11.8$ns; $2\sigma) = 4.9$ of surviving $^{26}$O 
ions and the corresponding time 
of flight, we obtain an upper limit for the 
lifetime of 5.7 ns at a 95\% confidence level. The upper limits for the energy of the
state and for its lifetime are shown in Fig.~\ref{fig:grig}, delimiting
the shaded area as the allowed region, which overlaps
with the calculated values for a pure $d^2$-state. A more complex
$^{26}$O ground-state configuration, however, cannot be excluded.

\begin{table}
\caption{Compilation of experimental and theoretical
results obtained in this work compared to previously published
results. All energies and lifetimes are given in keV and ns,
respectively. Energies are measured from the
g.s. energy of $^{24}$O. \label{tab}}
\begin{tabular}{ c c c c c }
\hline
\hline
& $E_r$ & $\Gamma$ & $\tau$ & Ref. \\ \hline
$^{25}$O(g.s.) & $725^{+54}_{-29}$ & $20^{+60}_{-20}$ & 
$ \geq 8.2 $ $\times$ $10^{-12}$ & this work \\
& $770^{+20}_{-10}$ & $172^{+30}_{-30}$ & - & \cite{hoff08} \\
average & $768 ^{+19}_{-9}$ & $160 ^{+30}_{-30}$ & $ 4.1^{+0.9}_{-0.6} \times 10^{-12}$ &\\
\hline
& 742 & \multicolumn{2}{l}{\hspace*{5mm}NN+3N + residual 3N} & this work \\
& 1301/1303 & \multicolumn{2}{l}{\hspace*{5mm}USDA/B} & \cite{brow06} \\
& 1002 & - & - & \cite{voly06} \\
\hline
$^{26}$O(g.s.) & $\leqslant$$40/120$\,\footnote{68\%/95\% c.l.} & $\geqslant$$1.2\times10^{-10}$\,\footnote{from
lifetime estimate, see text and 
Fig.~\ref{fig:grig}} & $\leqslant$$5.7$\,\footnote{95\% c.l.} & this
work \\
& $150^{+50}_{-150}$ & - & - & \cite{lund12} \\
\hline
& 40 & \multicolumn{2}{l}{\hspace*{5mm}NN+3N + residual 3N} & this work \\
& 501/356 & \multicolumn{2}{l}{\hspace*{5mm}USDA/B} & \cite{brow06} \\
& 21 & 0.02 & - & \cite{voly06} \\  
\hline
$^{26}$O(e.s.) & $4225^{+227}_{-176}$ & & & this work \\ 
\hline 
\hline
\end{tabular}
\end{table}

It would therefore be very interesting to determine the lifetime
experimentally more precisely in order to gain insights on the
structure of $^{26}$O. Such a measurement would be rather
straightforward using a similar method as described here, but placing
the target directly in front of the dipole. The decay curve of
$^{26}$O would translate into a fragment-mass distribution
depending on the decay position. In addition, the neutrons will be
detected not only at zero degree, but also in the bending direction,
again directly reflecting the decay curve. With the intensities
available, \textit{e.g.}, at the RIBF facility at RIKEN, a precise value could
be obtained from such an experiment with the SAMURAI setup.

\subsection{Comparison with shell-model calculations}

\begin{figure}
\includegraphics[clip=,keepaspectratio,width=1.0\columnwidth]{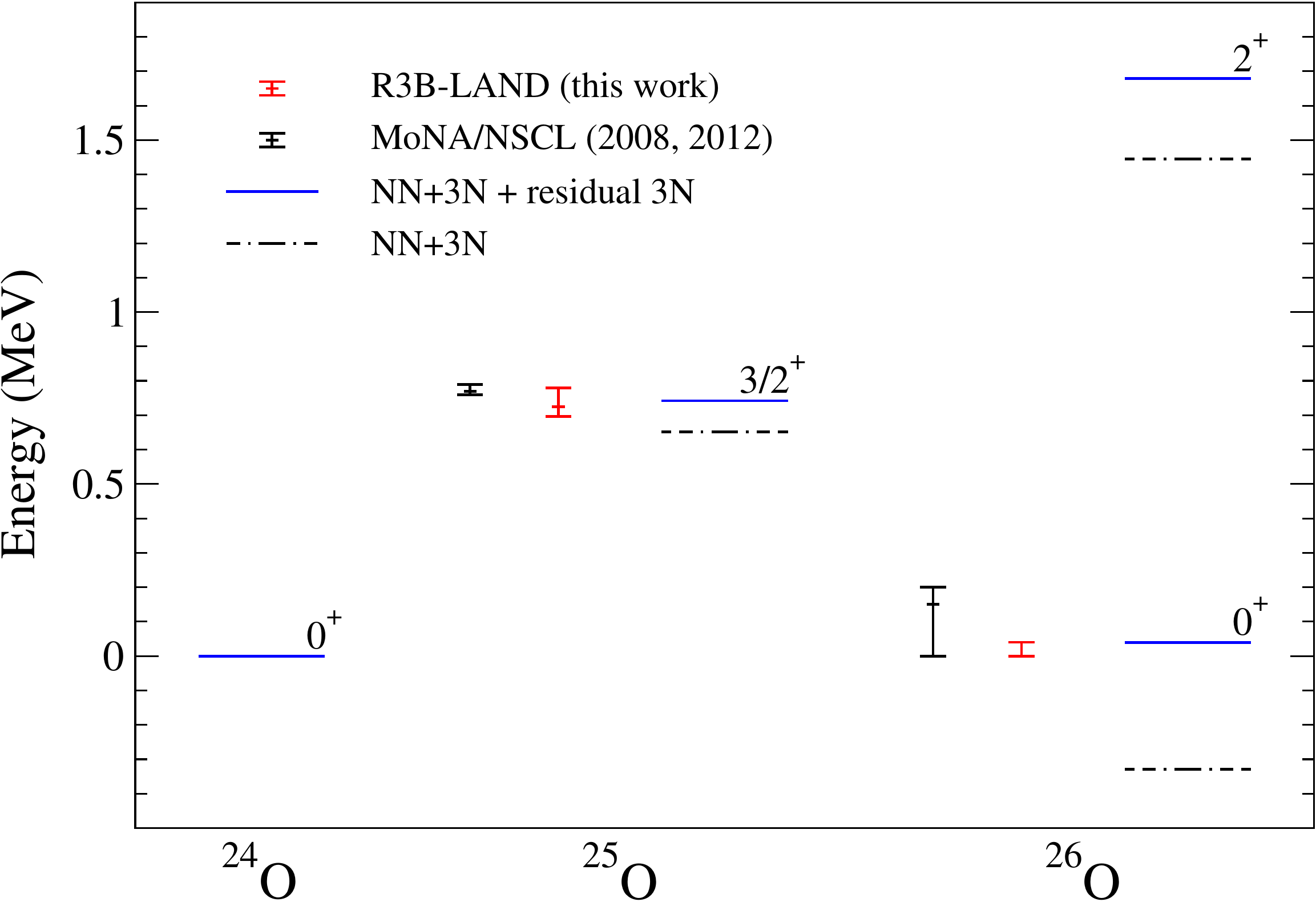}
\caption{\label{fig:theo} (color online) Comparison of the experimental
$^{25}$O and $^{26}$O energies with theoretical shell-model
calculations based on chiral NN and 3N forces (NN+3N) and including
residual 3N forces. Note that the contribution from residual 3N forces is 0.1 MeV 
in $^{24}$O. The data labeled as MoNA/NSCL are from Refs.~\cite{hoff08,lund12}.}
\end{figure}

We compare the ground-state energies of $^{25,26}$O to theoretical
shell-model calculations based on chiral effective field theory
potentials combined with renormalization-group methods to evolve
nuclear forces to low-momentum interactions~\cite{PPNP}. Our results are based
on chiral NN and 3N forces, where the single-particle energies and
two-body interactions of valence neutrons on top of a $^{16}$O core
are calculated following Refs.~\cite{holt11,gallant12} without
adjustments. Fig.~\ref{fig:theo} shows the predicted ground-state
energies obtained by full diagonalization in the valence shell
$H_{\mathrm{NN+3N,2b}}|\Psi_{\mathrm{NN+3N,2b}}\rangle=E_{\mathrm{NN+3N,2b}}|\Psi_{\mathrm{NN+3N,2b}}\rangle$. 
In addition to the 3N contribution to single-particle energies and
two-body interactions of valence neutrons, obtained by normal-ordering,
a third contribution is given by the weaker residual three-valence-neutron forces.
Here, we focus on the relative contribution from these residual forces,
which become more important with increasing neutron number along isotopic chains~\cite{Fermi}.
In order to quantify the relative contribution from
residual 3N forces, we use the wavefunction $|\Psi_{\mathrm{NN+3N,2b}}\rangle$
and calculate the correction
$\Delta E_{\mathrm{3N,res}}=\langle\Psi_{\mathrm{NN+3N,2b}}|V_{\mathrm{3N,res}}|\Psi_{\mathrm{NN+3N,2b}}\rangle$.
The repulsive contribution increases from 0.1 to 0.4~MeV for $^{24-26}$O,
highlighted by the arrows in Fig.~\ref{fig:theo}. Because the ground states of
$^{25,26}$O are very narrow, thus quasi-bound, they can be treated
fairly well in a bound-state approximation. The remarkable agreement
with experiment should, however, be considered with caution, as the continuum
needs to be included. The expected contribution is about 200~keV to both
$^{25,26}$O~\cite{hagen12}, so relative energy differences will be smaller. 
We have also explored uncertainties in the calculation of the single-particle 
energies and valence interactions, which would increase the ground-state 
energy relative to $^{24}$O for $^{25,26}$O by $0.2-0.3$~MeV.
In Table~\ref{tab}, we also compare with the phenomenological USDA/B
interactions~\cite{brow06}, which predict too high energies for
$^{25,26}$O. Better agreement is found in Table~\ref{tab} for the
continuum shell model~\cite{voly06} and in recent coupled-cluster calculations
with 0.4~MeV and 0.1~MeV for $^{25}$O and $^{26}$O, respectively~\cite{hagen12}.

\subsection{$^{26}$O excited state}

We now turn back to the group of events around 4~MeV in Fig.~\ref{fig:26O}, which we interpret as resulting
from the population of an excited state in $^{26}$O. 

Assuming first that all events above the
ground state could be explained by a non-resonant background (Eq.~\ref{eq_background}), this 
yields a fit to the data in the 1 to 8~MeV range as shown in Fig.~\ref{fig:26O} (bottom) 
by the blue dotted curve. 
The probability that the observed number of counts in the 4--5 MeV bin are explained by this
fit --- yielding an expectation value of 1.60 --- is less than 0.018. In turn, the probability that the 
accumulation of events in this energy region corresponds to a peak is larger than 98\% 
($\approx 2.5 \sigma $ significance)  even in a worst case scenario.

The use of a more realistic description of the data which includes three contributions
(one from the $^{26}$O ground state resonance, another from an excited state around 4 MeV, 
and a third contribution for the background) results in the fit to the data as displayed
by the black solid line in Fig.~\ref{fig:26O} (bottom), which represents
the fit function integrated over the experimental bin width.
The contribution of the background to the total fit is shown by the red dashed line, resulting
in a probability of 1.2$\times 10^{-6}$ 
for the events between 4 and 5~MeV to belong to the background,
which is equivalent to a significance of $4.85 \sigma $ for a peak structure. 

The theoretical calculations based on chiral NN and 3N forces predict
a first excited $2^+$ state in $^{26}$O at 1.6~MeV above the $^{26}$O ground state.
It is found at 1.9/2.1~MeV for USDA/B interactions and at 1.8~MeV for Ref.~\cite{voly06}.
Experimentally, the events at 4~MeV in the three-body
energy spectrum (Fig.~\ref{fig:26O}) provide an indication for
an excited state in $^{26}$O, with a most probable energy of
$4225^{+227}_{-176}$~keV. As for the ground-state resonance,
the response functions
have been used to approximate the line shape.
The minimum $\chi^{2}$ corresponds to 
the energy bin from 4200 to 4250~keV, which is shown in Fig.~\ref{fig:26O} including the experimental response. 
The errors have been determined from the  $\chi^{2}$-distribution using the interval given by $\chi^{2}_{min}+1$.

A likely candidate would be a proton-hole
state populated after knockout from the $^{27}$F $0p_{1/2}$ shell
(rather than from the $0d_{5/2}$ valence orbital). In order to investigate
proton (and neutron) cross-shell excitations, we have carried out
$(0+1)\hbar\omega$ calculations in the $spsdpf$ space using the WBP
interaction~\cite{WBP}, for which the first and second $2^+$ energies
are located at 2.3 and 5.4~MeV, respectively. The first state with a proton excitation
component from $0p_{1/2}$ to $0d_{5/2}$ is a $3^-$ state at a higher
energy of 5.4~MeV. Its proton contribution is also mixed with neutron
excitations and considerably weaker than for the corresponding $3^-$
state at 5.0~MeV in $^{18,20}$O. Note that $^{26}$O is bound by 1.0~MeV for the
WBP interaction, such that 1~MeV uncertainties are possible and the continuum should be included.
The lowest negative parity states predicted by the WBP interaction are a quartet of $3^-, 2^-, 1^-, 0^-$
neutron excitations from $0d_{3/2}$ to $1p_{3/2}$ at 3.7, 4.1, 4.5,
4.9~MeV (with centroid at 4.1~MeV). A conclusion on the character of the 
resonance-like structure 
around 4~MeV in the experimental spectrum cannot be drawn from the present study. 
A high-statistics experiment, which would allow for an investigation of the correlations in
the three-body decay, is necessary to shed light on the structure and quantum numbers
of this state.

\section{comment on the $^{26}$O lifetime}

During the final stages of the refereeing process two Letters have been published in PRL discussing
the lifetime of $^{26}$O, one experimental \cite{kohl13} and the other theoretical \cite{grig13}. In the experimental
work by the NSCL-MoNA group, a half-life  $T_{1/2} = 4.5_{-1.5}^{+1.1}$(stat)$\pm3$(syst)~ps has been extracted, 
corresponding to a lifetime of 6.5~ps. At an 82\% confidence level, a finite lifetime of the $^{26}$O ground state is
claimed  \cite{kohl13}, suggesting the possibility of two-neutron radioactivity. Our upper 
limit of 5.7 ns (95\% c. l.) is in agreement with this finding. 
Even the combination of both results does not allow for a firm conclusion ($5\sigma$ signal) on the possibility of
neutron radioactivity.
It is therefore of utmost importance
to perform a dedicated and optimized experiment with good statistics to measure the lifetime of $^{26}$O. 
With the method we have proposed in this paper, namely to measure the decay of $^{26}$O in a magnetic dipole field, it
will be difficult, however, to measure a lifetime shorter than 10--100~ps. The sensitivity range depends on the field strength
and in particular on the angular resolution for the neutron detection, limiting the sensitivity with present detectors 
to around 100~ps, and to around 10~ps at the future R3B facility at FAIR. A more elaborated discussion of the method
proposed in section III.B of this paper has been published in the meantime by Thoennessen \textit{et al.}~\cite{thoe13}, together
with a more detailed discussion of the method used by Kohley \textit{et al.}~\cite{kohl13}. 

The new theoretical estimate 
by Grigorenko \textit{et al.}~\cite{grig13} implies a very low upper limit for the resonance energy of the $^{26}$O ground state 
of around 1~keV \cite{grig13}, which is in agreement with the upper limit of 40/120~keV (68\%/95\% c. l.) as derived
in the present work. 

\section{CONCLUSION}

In summary, we have investigated the
ground-state energy, width, and lifetime of the unbound oxygen
isotopes $^{25}$O and $^{26}$O. Our results are in very good agreement
with theoretical shell-model calculations based on chiral NN and 3N
forces, where the ground-state energy of these extremely neutron-rich
isotopes becomes increasingly sensitive to 3N forces among the valence
neutrons. The $^{26}$O ground state is unbound by less than 120~keV,
and our measurement provides a limit on the lifetime of $\leqslant$5.7~ns (both at 95\% c.l.). 
We also
obtained indications for an excited state of $^{26}$O located at
about 4~MeV. Different possibilities for the nature of this state
exist, making it an exciting case for future calculations and
experiments with higher statistics.

\begin{acknowledgments}
We thank B.\ A.\ Brown for helpful discussions on the WBP interaction.
This work was supported by the Helmholtz International Center for 
FAIR within the framework of the LOEWE program launched by the state
of Hesse, by the Helmholtz Alliance Program of the Helmholtz 
Association, contract HA216/EMMI ``Extremes of Density and
Temperature: Cosmic Matter in the Laboratory'', by the GSI-TU
Darmstadt Cooperation agreement, by the BMBF under Contracts
No. 06DA70471, 06DA9040I, 06MT238, by the DFG cluster of excellence 
Origin and Structure of the Universe, by the US DOE Grants DE-FC02-07ER41457, 
DE-FG02-96ER40963, via the GSI-RuG/KVI collaboration agreement,
and by the Portuguese FCT, project PTDC/FIS/103902/2008.
\end{acknowledgments}

\end{document}